# DroidMark – A Tool for Android Malware Detectionusing Taint Analysis and Bayesian Network

Dhruv Rathi
Department of Computer Engineering
Delhi Technological University
New Delhi, India
e-mail: dhruvrathi15@gmail.com

Dr Rajni Jindal
Head of Department, Department of Computer Engineering
Delhi Technological University
New Delhi, India
e-mail: rajnijindal@dce.ac.in

*Abstract*—With the increasing user-base of Android devices & advents of technologies such as Internet Banking, delicate user data is prone to be misused by malware and spyware applications. As the app-developer community increases, the quality reassurance could not be justified for every application and a possibility of data leakage arises. In this research, with the aim to ensure the application authenticity, Deep Learning methods and Taint Analysis are deployed on the applications. The detection system named DroidMark looks for possible sinks and sources of data leakage in the application by modelling Android's lifecycles and callbacks, which is done by Reverse Engineering the APK, further monitoring the suspected processes and collecting data in different states of the application. DroidMark is thus designed to extract features from the applications which are fed to a trained Bayesian Network for classification of Malicious and Regular applications. The results indicate a high accuracy of 96.87% and an error rate of 3.13% in detection of Malware in Android devices.

__________________________________________________*\*\*\*\*\**__________________________________________________

## I. INTRODUCTION

Mobile devices with easy access and continuous development on optimizing the system to support multi-functionalities and running high-level applications on relatively lower computationally powered devices have led to high advancements in the devices. Android devices, in particular, are the most popular Operating System(OS) among all with 352 million Smartphones running the OS, which accounts for 81.7% of total industry. The primary reason for the popularity of the Google-owned OS is the huge Open-Source community that is continuously involved in the development of the system. The OS architecture has Linux at the kernel, hence the developers are free to modify the Linux kernel to fit their needs. The ubiquity of the Android Platform has not gone unnoticed by Malware and Spyware developers. Android devices have been targeted in the past by Malicious applications and widespread attacks targeting mass users have been exploited. A malicious application targets the device with the aim to collect user information which otherwise should be kept discreet. Some examples could be credit card information or sharing of location when not needed, sending and intercepting SMS and voice calls, etc.

A plethora of attempts have been made in the past to target the mass user, DroidDream (Android.Rootcager)[1] in 2011, infected 60 different wide used apps, and breached the Android Security sandbox, installed additional malicious software and stealing of data. Android.PjappsM[2], a 2010 botnet infected huge number of devices by that would then launch attacks on websites to steal additional data and infect more devices, initiating a chain reaction.

Malware detection can be modelled using different approaches, as shown in Figure 1[2].

A Signature-based detection technique requires some previous information of the features which are known to be malicious to categorize the program under inspection. In contrast, the Anomaly-based technique uses its knowledge of what is normal to decide the maliciousness of a program under inspection.

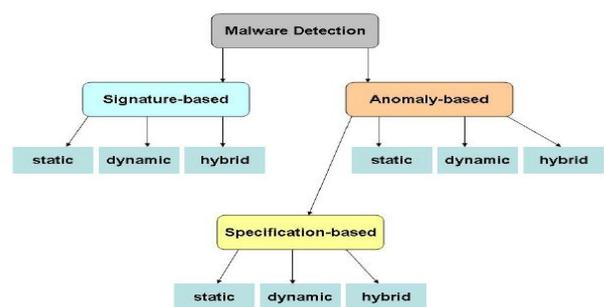

Fig 1. Classification of Malware detection techniques.

The presented research exploits the fact that a malicious application necessarily makes API calls and executes processes which when tested using proper tools proves to be out of context. The research presents a tool DroidMark which maps the context-flow, object-sensitivity, and lifecycle-awareness which results in potential sinks of data leakage, the sinks are tracked by deploying optimized process tracing techniques. The data is collected using various mobile sources, maintained on an allocated server, further data cleaning techniques are

71

employed. The final data of the suspicious processes is obtained and passed through a Deep Learning model of trained Bayesian Network which classifies the application in Regular/Malicious.

## II. RELATED WORKS

Utilizing Deep Learning for Android Malware detection has not been explored to a great depth yet, following we have discussed some related works,

Zhenlong Yuan et. al [1] attempts to classify an Android application into Malicious and Normal using Support Vector Machine, a Deep Learning algorithm, achieving a 96.76% detection accuracy on a set of 20,000 Android applications.

Xudong He et. al[2] employs High-Level Petri Nets for modelling and analyzing the Android Permissions Framework and exploring Malicious leakage at the permissions level.

Joshua Abah et. al[3] monitors different API and employs Machine Learning algorithm K Nearest Neighbors to classify the data into Normal or Benign using WEKA tool, reporting an accuracy of 93.75%.

## III. EXPERIMENTAL AND COMPUTATIONAL DETAILS

### A. Layout

The flow of the tool DroidMark, developed in the research process can be best explained by visualizing it in 3 steps. The first step is called the Taint Analysis. The second step is responsible for the collection of data. The final step is the Deployment of Deep Learning of models to classify the data for detection of Malicious applications.

### B. What is Taint Analysis?

Taint Analysis analyzes the application and presents potentially malicious data flow to human analysts or to automated malware detection tools.

*FlowDroid:* FlowDroid is an open-source tool for static-taint analysis which is integrated into DroidMark to detect sensitive data flows in Android applications. FlowDroid is a full context, flow, field, object sensitive tool with the aim to correctly model the Android lifecycle, UI interaction, system-event handling by analyzing the application's byte-code and configurational files to find potential privacy leaks, either created by mistake or with malicious intentions. In Android app's code there is no Main function unlike normal JAVA code, thus it becomes difficult to obtain the control flow graph simply by reviewing the entry and exit points of the classes, thus as an effective way around the life-cycle functions such as onCreate(), onStart(), onDestroy(), etc. contains various callbacks which notify the application flow, the system events and GUI events.

Targeting the tracking of sensitive data flows and potential Sources and Sinks of data leakage in the Android application, FlowDroid searches for lifecycle and callback methods as well as calls to potential data leakage points. FlowDroid formalized the data flow analysis to a taint analysis on IFDS framework. In short, FlowDroid generates a report which indicates the potential data leakage Sources and Sinks by different API calls and process calls.

*Malware used:* For the current explanation, the Android-virus Elite[4] is taken, Elite virus forces the following malware into the device:
- Send SMS continuously from the device till the balance is nil.
- Block SMS messenger and related applications.
- Run an unauthorized background process, with a handler on Boot Receiver.
- Wipe out SD-card data completely.
- Retrieve data input by the user in another application.

The dynamic usability of DroidMark allows the system to enter the APK and FlowDroid follows two-step process of analysis consisting of a forward-taint propagation which tracks the tainted value, and a backwards-alias that resolves to alias in the heap model.

## IV. DATA LEAKAGE DETECTION: SOURCES AND SINKS

The report generated by FlowDroid using the original features extracted using the data flows and other components contains the list of Source and Sink method names thus generating a feature vector suspicious data leakage points:

```
returnLeakage(application.APK)
{
    result = [(source1_method -> sink1),
        (source1_method -> sink2)
        .
        .
        (sourceM_method -> sinkN_method)]

    return result
}
```

If data flow exists between a source method and a sink method, the feature vector is updated with a value of 1, else it is assigned 0. There are different sources of these sinks such as external libraries which may not be interacting with the application at any points in the lifecycle, thus the feature vector will remain too sparse to produce good results as the number of susceptible sinks are higher than those getting involved.

### A. SuSI

To overcome this setback, the SuSI method as discussed in Dil Zhu et. al [4] for enhanced automation of sources and sinks detection from the source code of an Android API. SuSI methodology proposes a categorization into 17 sources and 19 sinks.

| S.No. | Source Category | Sink Category |
|---|---|---|
| 1 | LOCATION_INFORMATION | PHONE_CONNECTION |
| 2 | NETWORK_INFORMATION | PHONE_CONNECTION |
| 3 | FILE_INFORMATION | EMAIL |
| 4 | BLUETOOTH_INFORMATION | BLUETOOTH |
| 5 | EMAIL | AUDIO |
| 6 | UNIQUE_IDENTIFIER | LOCATTON_INFORMATION |
| 7 | ACCOUNT_INFORMATION | PHONE_STATE |
| 8 | SYNCHRONIZATION_DATA | SYNCHRONIZATION_DATA |
| 9 | SMS_MMS | NETWORK |
| 10 | SYSTEM_SETTING | SMS_MMS |
| 11 | CONTACT_INFORMATION | FILE |
| 12 | CALENDAR_INFORMATION | LOG |
| 13 | IMAGE | CONTACT_INFORMATION |
| 14 | BROWSER_INFORMATION | CALENDAR_INFORMATION |
| 15 | NFC | SYSTEM_SETTING |
| 16 | DATABASE_INFORMATION | SYNCHRONIZATION_DATA |
| 17 | NO_CATEGORY | NFC |
| 18 | | BROWSER_INFORMATION |
| 19 | | NO_CATEGORY |

Table 1. SuSI categories of sensitive sources and sinks

Providing Elite.APK as the input file to a modified SuSI generates the following outputs: which are suspected to cause data leakage in the malicious application.

Process List: {
**'com.samsung.ui',**
**'datapole.rathi.monitor',**
**'com.elite.AlarmReceiver',**
**'com.elite.SMSReceiver', 'com.android.bluetooth',**
**'android.telephony.SMSManager',**
**'com.sec.imsservice',**
**'com.elite.BootReceiver'**}

*B. Data Collection by Monitoring Sources and Sinks*

DroidMark has provided us with a list of suspected processes with the help of FlowDroid and SuSI frameworks.

The next step is the real-time data collection from devices. The monitoring schema is designed based on the fact that a malicious application requires user interaction to activate its functionality on the target device. The analysis of Elite.APK has provided us with, and thus 4 different system process calls, in addition, 1 more feature is added to prepare the feature vectors for the Bayesian Network as indicated in Table-1.

| **android.telephony.SmsManager** |
|---|
| **com.elite.BootReceiver** |
| **com.elite.SMSReceiver** |
| **com.elite.AlarmReceiver** |
| **Screen Wake State** |

Table 2. Monitored processes by DroidMark

Describing the features in Table-2,

A feature table for further reference is created,

**feature[] = {**android.telephony.SmsManager,
   com.elite.BootReceiver,
   com.elite.SMSReciever,
   com.elite.AlarmReceiver,
   ScreenWakeState **},**

feature[1] controls the incoming and outgoing of SMS services, which is an OS level call, feature[2], feature[3] and feature[4] are packages in Elite.APK which receive SMS, Alarm state Listeners respectively.

feature[4] also logs in whether the device state is of Deep Sleep or is Awake during the Calling and Callback to the listeners of the above processes vector.

A foreground service is created by DroidMark which operates to completely focus on the process in feature[] array, this is done by creating a function killAllProcess(), this function is designed to kill all the running processes which are marked suspected by FlowDroid, which in the present case of Elite.APK are the contents of the array 'feature[]'.

This gives DroidMark, the advantage that the monitoring of features such as SMS Manager will be of the application under process and not that of other parallel running applications.

| feature[1] | feature[2] | feature[3] | Wake/Sleep | Type |
|---|---|---|---|---|
| 1 | 0 | 1 | 0 | Malicious |
| 1 | 0 | 1 | 1 | Normal |

Table 3. Classification Example

Considering Table-3, even though the device reports the process of outgoing SMS via the process call android.telephony.SmsManager, this is considered Normal when the device is in awake state(state 1), but is considered Malicious when the device undergoes Deep Sleep (state 1), but is considered Malicious when the device undergoes Deep Sleep (state 0), because the act of sending SMS requires an active user interaction.

Similar feature vectors are parsed and converted into Attribute Relation File Format (ARFF) with the process name and the states appended to each instance and further used as input to train a Bayesian Network.

*C. Bayesian Networks*

Bayesian networks are a type of Probabilistic Graphical Model. Each node in the graph represents probabilistic dependencies among the corresponding random variables. A

73

Bayesian network B is an annotated acyclic graph that represents a Joint Probability Distribution(JPD) over a set of random variables **V**.

The network is defined by a pair B = <G, Θ>, where G is a Direct Acyclic Graph whose nodes $X_1, X_2,...,X_n$ represents random variables, and whose edges represent the direct dependencies between these variables. The graph G encodes independence assumptions, by which each variable Xi is independent of its non-descendants given its parents in G. The second component Θ denotes the set of parameters of the network. This set contains the parameter $\theta_{x_i | \pi_i} = P_B(x_i|\pi_i)$ for each realization $x_i$ of $X_i$ conditioned on $\pi_i$, the set of parents of Xi in G. Accordingly, B defines a unique Joint Probability Distribution over V, namely:

$$P_B(X_1, X_2,...,X_n) = \prod_{i=1}^{n} P_B(X_i | \pi_i) = \prod_{i=1}^{n} \theta_{X_i | \pi_i}$$

*D. WEKA*

Waikato Environment for Knowledge Analysis(WEKA) is an open-source library in JAVA that contains set of algorithms that can be easily implemented in an Android device with a specialized Android application. The Weka.jar file is added as an external library to the Android studio project, into which an appropriate GUI is created to invoke various features programmatically using a set of Java APIs.

## V. RESULTS AND DISCUSSION

*A. Feeding Data to WEKA*

The data collected using different users of the malicious application Elite.APK is collected on a centred Head device which has the DroidMark set-up installed. The data collected is to be fed to the Bayesian Network to complete the Bayesian Network training stage.

The data collected from multiple sources need to be pre-processed in the Attribute Relation File Format (ARFF) format which is the requirement of WEKA.

Sample data collected in ARFF format is explained in Figure 2.

Each attribute has the value 0 or 1, which is decided by whether the system process/ system API is called in the current state of the system by different application packages. The Class attribute is the unknown which needs to be predicted and thus is represented by '?', which is the standard format for unknown variables in ARFF. The @data contains the processes of different applications and data about whether there is a leakage simultaneously activation of any of the data leakage processes, represented by 0/1.

```
@relation RunningProcessVectors
@attribute ProcessName {'com.samsung.ui','datapole.rathi.monitor',
                        'com.elite.AlarmReceiver',
                        'com.elite.SMSReceiver',
                        'com.android.bluetooth',
                        'android.telephony.SMSManager',
                        'com.sec.imsservice','com.elite.BootReceiver'}
@attribute BootReceiver {0,1}
@attribute SMSReceiver {0,1}
@attribute AlarmReceiver {0,1}
@attribute android.telephony.SmsManager {0,1}
@attribute ScreenWake {0,1}
@attribute Class {Regular, Malicious}
@data
'com.samsung.ui',0,0,0,0,1,?
'datapole.rathi.monitor',1,0,1,0,1,?
'com.elite.AlarmReceiver',0,1,1,0,1,?
'com.elite.SMSReceiver',1,1,0,1,1,?
'com.android.bluetooth',0,0,0,0,1,?
'android.telephony.SMSManager',0,1,0,1,1,?
'com.sec.imsservice',0,0,0,0,1,?
'com.elite.BootReceiver',1,1,0,1,1,?
'com.samsung.ui',1,0,1,0,0,?
'datapole.rathi.monitor',1,0,1,0,0,?
'com.elite.AlarmReceiver',1,1,1,0,0,?
'com.elite.SMSReceiver',0,1,0,1,0,?
'com.android.bluetooth',0,0,0,0,0,?
'android.telephony.SMSManager',1,1,0,1,0,?
'com.sec.imsservice',0,0,0,0,0,?
'com.elite.BootReceiver',1,1,1,1,0,?
'com.samsung.ui',1,0,1,0,0,?
'datapole.rathi.monitor',1,0,1,0,0,?
'com.elite.AlarmReceiver',1,1,1,1,0,?
'com.elite.SMSReceiver',1,1,0,1,0,?
'com.android.bluetooth',0,0,0,0,0,?
'android.telephony.SMSManager',1,1,0,1,0,?
```

Fig2. eliteDATA.arff – Parsed data sample in ARFF format

*B. The Criterion for Accuracy of Results*

The output gives us several combinations of criterions to find the accuracy, similar approaches have been used earlier by [13]. In our context, the relevant measures are discussed below,

1. **TPR**:- Rate of True positives (TP), i.e., instances which are correctly classified as a class.
2. **FPR**:- Rate of False Positive (FP), i.e. instances which are falsely classified as a class.
3. **Precision**:- The proportion of instances that are truthful of a class divided by the total instances of that class.
4. **Recall**:- The proportion of instances classified as a given class divided by the actual total in that class.
5. **ROC:-** ROC value should be approaching 1 for an efficient classifier.

*C. Output and Accuracy*

Similar data as in Figure 2 is fed in the Bayesian network for the purpose of classifying the application in Malicious/Normal category.

```
=== Run information ===

Scheme:       weka.classifiers.bayes.BayesNet -D
Relation:     RunningProcessVectors
Instances:    32
Attributes:   7
              ProcessName
              BootReceiver
              SMSReceiver
              AlarmReceiver
              android.telephony.SmsManager
              ScreenWake
              Class
Test mode:    10-fold cross-validation

=== Classifier model (full training set) ===

Bayes Network Classifier
not using ADTree
#attributes=7 #classindex=2
Network structure (nodes followed by parents)
ProcessName(8): SMSReceiver
BootReceiver(2): SMSReceiver
SMSReceiver(2):
AlarmReceiver(2): SMSReceiver
android.telephony.SmsManager(2): SMSReceiver
ScreenWake(2): SMSReceiver
Class(2): SMSReceiver
LogScore Bayes: -172.44387603519843
LogScore BDeu: -218.1484341583698
LogScore MDL: -217.64854929907725
LogScore ENTROPY: -174.32685051408063
LogScore AIC: -199.32685051408063
```

Fig 3. Model structure and SuSI output

```
=== Stratified cross-validation ===
=== Summary ===

Correctly Classified Instances        31        96.875 %
Incorrectly Classified Instances       1         3.125 %
Kappa statistic                        0.9375
Mean absolute error                    0.0678
Root mean squared error                0.1609
Relative absolute error               13.4539 %
Root relative squared error           31.9165 %
Total Number of Instances             32
```

Fig 4. Result and Accuracy

```
=== Confusion Matrix ===

 a  b   <-- classified as
16  0 | a = 0
 1 15 | b = 1
```

Fig 5. Confusion Matrix

```
=== Detailed Accuracy By Class ===

               TP Rate  FP Rate  Precision  Recall  F-Measure  MCC    ROC Area  PRC Area  Class
               1.000    0.063    0.941      1.000   0.970      0.939  1.000     1.000     0
               0.938    0.000    1.000      0.938   0.968      0.939  1.000     1.000     1
Weighted Avg.  0.969    0.031    0.971      0.969   0.969      0.939  1.000     1.000
```

Fig 6. Summarized Text Results

*D. Discussion on Results*

From the outputs shown in 3.3, it can be easily inferred that the Bayesian Network provides a very high accuracy, precisely, DroidMark correctly classifies 96.88% of the samples with error percentage of 3.12% as shown in the output of Figure 5, and summarized in Figure 6 in accordance with Criterion for Accuracy as set in 3.2

Discussing the results further in detail in the context of 3.2, the Precision of Malicious and Regular applications is 0.941 and 1.000, and the TPR (which equals the Recall) are 0.938 and 1.000 respectively.

The confusion matrix from Figure 5 shows the correctly classified samples and the misclassified malicious samples from the experiment, it is evident that the malicious class samples classified as being of the Normal class were responsible for the incorrect classification.

An accuracy of 96.88% achieved by the Bayesian Network classifier indicates a significant improvement in detecting a Malicious application, moreover, the time taken to process the collected data is found to be in range of 0.5-10 seconds, thus making the model a good choice for integration in real time devices.

CONCLUSION

In the proposed work, Taint Analysis and Machine Learning approach has been followed to implement an efficient Malware detection system for Android devices. Taint Analysis attempts to construct the life-cycle by Reverse Engineering the APK file of the application under review, this step is performed to find potential sources of sensitive Data Leakage without the consent of the user. Thus, as an output, various processes are marked as Sources and Sinks of leakage. This is done with a modified version of DroidFlow and SuSI frameworks. These processes are monitored for the purpose of data collection via a specialized Android application created with JAVA. WEKA tools were used to train and test a Bayesian Network inside the

75

Android system by feeding the data collected. The results with an accuracy of 96.88% and an error rate of 3.22%, clearly indicate that DroidMark promises an efficient system in a real-time implementation in Android devices,